\PassOptionsToPackage{table}{xcolor} %
\documentclass[sigconf]{acmart}
\usepackage[utf8]{inputenc}
\usepackage[T1]{fontenc}
\usepackage[table]{xcolor}
\usepackage{hyperref}
\usepackage{url}
\usepackage{tabularx}
\usepackage{booktabs}
\usepackage{amsfonts}
\usepackage{amsmath}
\usepackage{nicefrac}
\usepackage{microtype}
\usepackage{balance}
\usepackage{pifont}
\usepackage{xspace}
\usepackage{graphicx}
\usepackage{tikz}
\usetikzlibrary{fit,positioning,calc,shapes,backgrounds}
\usepackage{pgfplots}
\usepackage{hyphenat}
\pgfplotsset{compat=1.14}
\usepgfplotslibrary{fillbetween}
\usepackage{wrapfig}

\usepackage{bm}
\usepackage[bbgreekl]{mathbbol}
\usepackage{amsfonts}

\DeclareMathOperator{\tr}{tr}

\newcommand{\mpara}[1]{\medskip\noindent{\bf #1}}

\definecolor{cycle1}{RGB}{228, 26, 28}
\definecolor{cycle2}{RGB}{55, 126, 184}
\definecolor{cycle3}{RGB}{77, 175, 74}
\definecolor{cycle4}{RGB}{152, 78, 163}
\definecolor{cycle5}{RGB}{255, 127, 0}
\definecolor{cycle6}{RGB}{153, 153, 153}%
\definecolor{cycle7}{RGB}{166, 86, 40}
\definecolor{cycle8}{RGB}{247, 129, 191}

\newcommand*{\bigO}{\mathcal{O}}

\newcommand{\cmark}{\textcolor{cycle3}{\ding{52}}} %
\newcommand{\xmark}{\textcolor{cycle1}{\ding{56}}}

\newcommand{\gnp}{Erd\H{o}s-R\'einy\xspace}

\newcommand{\thiswork}{\textsc{SGR}\xspace}
\newcommand{\ours}{\thiswork} %
\newcommand{\sgr}{\textsc{SGR}}

\newcommand{\netsimile}{\textsc{NetSimile}\xspace}
\newcommand{\fgsd}{\textsc{FGSD}\xspace}
\newcommand{\spg}{\textsc{SP}\xspace}
\newcommand{\gk}{\textsc{GK-4}\xspace}
\newcommand{\wl}{\textsc{WL}\xspace}
\newcommand{\mlg}{\textsc{MLG}\xspace}

\newcommand*{\graphpair}[2]{(#1, #2)}

\newcommand{\bb}[1]{\bm{\mathrm{#1}}}

\newcommand{\vertices}{V}

\newcommand{\edges}{E}
\newcommand{\graphcol}{\mathcal{G}}
\newcommand{\graph}{G}
\newcommand{\graphexp}[1]{\graph_{#1} = \graphpair{\vertices_{#1}}{\edges_{#1}}}
\newcommand{\sig}{\bb{\sigma}}

\newcommand{\diam}{\mathbf{D}}

\newcommand{\adj}{\bb{A}}
\newcommand{\lapl}{\bb{L}}
\newcommand{\hk}{\bb{H}_t}

\newcommand{\win}{\cellcolor{cycle3!30}}

\setcopyright{rightsretained}

\title{\ours: Self-Supervised Spectral Graph Representation Learning}

\setcopyright{rightsretained}

\acmConference[KDD'18 Deep Learning Day]{ACM KDD'18 Deep Learning Day}{August 2018}{London, UK}
\acmYear{2018}
\copyrightyear{2018}
\copyrightyear{2018}

\begin{document}
\author{Anton Tsitsulin}
\affiliation{%
  \institution{Hasso Plattner Institute}
  \city{Potsdam}
  \country{Germany}
}
\author{Davide Mottin}
\affiliation{%
  \institution{Hasso Plattner Institute}
  \city{Potsdam}
  \country{Germany}
}
\author{Panagiotis Karras}
\affiliation{%
  \institution{Aarhus University}
  \city{Aarhus}
  \country{Denmark}
}
\author{Alexander Bronstein}
\affiliation{%
  \institution{Technion}
  \city{Haifa}
  \country{Israel}
}
\author{Emmanuel M\"uller}
\affiliation{%
  \institution{Hasso Plattner Institute}
  \city{Potsdam}
  \country{Germany}
}

\begin{abstract}

Representing a graph as a vector is a challenging task; ideally, the representation should be easily computable and conducive to efficient comparisons among graphs, tailored to the particular data and analytical task at hand. Unfortunately, a ``one-size-fits-all'' solution is unattainable, as different analytical tasks may require different attention to global or local graph features. We develop \ours, the first, to our knowledge, method for learning graph representations in a {\em self-supervised\/} manner. Grounded on spectral graph analysis, \ours seamlessly combines all aforementioned desirable properties. In extensive experiments, we show how our approach works on large graph collections, facilitates self-supervised representation learning across a variety of application domains, and performs competitively to state-of-the-art methods without re-training. \end{abstract}

\maketitle

\section{Introduction}\label{sec:introduction}

A multitude of data in various domains, from natural sciences to sociology, is represented as collections of graphs. For example, on a small scale molecules are modeled by atoms and their atomic bonds as nodes and edges in large graph collections. While on a larger scale collections of social networks are analyzed by their community structures within the networks. Analytical tasks run on such collections to classify, for instance, which drugs can be used for the treatment of a disease or how molecules cluster together in functional groups. To fully discern a graph's properties, representation learning for such analysis requires a {\em multiscale\/} view of a graph. Representations have incorporated properties ranging from {\em local\/} (e.g., atomic bonds) to {\em global\/} (e.g., community structures).

\emph{Kernel} methods~\cite{borgwardt2005, shervashidze2009, shervashidze2011, yanardag2015, kondor2016}, dominate the field of graph analytics, as they define functional similarities among pairs of graphs and can perform tasks such as graph classification. Among a plethora of graph kernels, to our knowledge, only the Multiscale Laplacian Graph kernel (MLG)~\cite{kondor2016} preserves multiscale properties. Still, such graph kernels require {\em direct comparisons} among pairs of graphs, hence scale quadratic in the size of any graph collection. As more and more data is stored as graph collections, such all-pairs direct comparisons methods are rendered inviable.

In another vein, \emph{graph representations} extract a vector signatures from a graph and perform comparison among those vectors in lieu of the graphs. Initial approaches derived such representations by hand-crafted statistics on the graph structure (e.g., the average node degree~\cite{berlingerio2013}).
Recently, supervised neural approaches for graph representation~\cite{atwood2016,niepert2016} attained competitive performance in supervised classification among graphs of a few tens of nodes. However, such neural methods are applicable to particular datasets only, as they require labels to be available; besides, they fail to scale to graphs of a few thousands of nodes. Most recently, Verma~et~al.~\cite{verma2017} proposed the {\em Family of Graph Spectral Distances\/} (FGSD): a graph representation method based on histograms of the biharmonic kernel. While FGSD representations are designed for classification tasks, these are neither expressive nor scalable enough to be used in both classification tasks.

In this paper we propose \ours, a self-supervised method for learning graph representations that is at the same time efficient to compute and customizable to multiple scales, analytical tasks, and datasets. \ours leverages a graph's Laplacian spectrum to learn a mapping a collection of graphs to their vector representation, by learning a simple single-layer perceptron on global structure recognition. The perceptron learns to distinguish synthetic graphs with community structure (i.e., sampled from a stochastic block model~\cite{karrer2011}) from random graphs by the \gnp model.
\ours representation is self-learning in the sense that it requires no real training data. We conduct an experimental study with several real datasets, using the ensuing graph signature representations on tasks such as graph classification by logistic regression. The results attest the superiority of our approach on classification tasks with real data.

\section{Related work}\label{sec:related-work}

Previous work on learning graph representations falls into three groups, outlined in Table~\ref{tbl:relatedwork}.

\begin{table*}[h]
\setlength{\tabcolsep}{2pt}
\setlength{\belowrulesep}{0pt}

\begin{center}
{
\small
\newcolumntype{C}{>{\centering\arraybackslash}X}
\begin{tabularx}{\textwidth}{Xp{3cm}CCCCC} %
\multicolumn{1}{C}{} & \multicolumn{1}{C}{} & \multicolumn{3}{c}{\textbf{Properties}} & \multicolumn{2}{c}{\textbf{Complexity}} \\
\cmidrule(lr){3-5}\cmidrule(lr){6-7}
\bf{Method}              & {\bf Type} & Learned &  Unsupervised   & Multi-scale    & Precomputation       & Comparison        \\ 
\midrule
SP~\cite{borgwardt2005}    & kernel & \xmark{} & \xmark{} & \xmark{} & $\bigO(1)$ &  $\bigO(n^3)$\\
WL~\cite{shervashidze2011} & kernel & \xmark{} & \xmark{} & \xmark{} & $\bigO(1)$ &  $\bigO(m \log(n))$\\
MLG~\cite{kondor2016}      & kernel & \xmark{} & \cmark{} & \cmark{} & $\bigO(k m + k^2 n)$ & $\bigO(k^3)$ \\
\midrule
PSCN~\cite{niepert2016}    & supervised NN & \cmark{} & \xmark{} & \xmark{} & $\bigO(dn\log(n))$ & $\bigO(dn\log(n))$ \\
DCNN~\cite{atwood2016}     & supervised NN & \cmark{} & \xmark{} & \xmark{} & $\bigO(dn^2)$ & $\bigO(dn^2)$ \\
\midrule
\netsimile{}~\cite{berlingerio2013}  & fixed representation & \xmark{} & \cmark{} & \xmark{} & $\bigO(n \log n)$ & $\bigO(d)$ \\
\fgsd{}~\cite{verma2017}   & fixed representation & \xmark{} & \cmark{} & \xmark{} & $\bigO(n^2)$ & $\bigO(d)$ \\
\midrule
\rowcolor{cycle2!10}
\textbf{\thiswork}         & self-learned representation & \cmark{} & \cmark{} & \cmark{} & $\bigO(k m + k^2 n)$ & $\bigO(d)$ \\
\end{tabularx}
}
\end{center}
\caption{Related work allowing graph comparison in terms of fulfilled (\cmark) and missing (\xmark) characteristics and complexity ($n$ nodes, $m$ edges, $k$ eigenvalues, $d$ representation dimensions).}\label{tbl:relatedwork}
\end{table*} %
\subsection{Kernel methods}

{\em Graph kernels\/}~\cite{gartner2003, borgwardt2005, shervashidze2011, yanardag2015, nikolentzos2017, kondor2016} are similarity functions among graphs, which perform an implicit transformation of graph structure to compare two graphs (e.g.\ Shortest-path (SP) kernel~\cite{shervashidze2011}). However, kernel methods are limited due to (i) high on-demand computational complexity at comparison time, which renders them inapplicable to large-scale graph comparisons, and (ii) rigidity: once a kernel is chosen, it cannot be tailored to the analytical task or dataset at hand.
The Multi-scale Laplacian Graph kernel (MLG)~\cite{kondor2016} is a mature work on this domain, as it adapts to different scales via an iterative information-propagation method. Yet it also raises a computational overhead cubic in Laplacian matrix eigenvalues.

\subsection{Supervised neural methods}

Advances in neural learning have led to the application of {\em supervised neural\/} approaches to classify collections of graphs. The Diffusion Convolutional Neural Network (\textsc{DCNN})~\cite{atwood2016} learns graph representations by averaging values after a diffusion process on a graph's nodes. 
Similarly, \textsc{Patchy-san}~\cite{niepert2016} learns a representation through a CNN filter after imposing a sampling order on nodes. 
Yet such approaches share the drawbacks of kernel methods: high computational overhead at comparison time and lack of variable adaptability to local or global structures.
Besides, the learning component in these neural methods is {\em supervised\/} by means of node and edge labels, raising an additional resource requirement.

\subsection{Fixed representation methods}

Another class of approaches eschew the supervised learning component of neural approaches. Such works started out using features engineered by aggregating local graph properties such as node degree and neighbors' degrees~\cite{bronstein2011a, berlingerio2013, bonner2016}.
However, in eschewing supervision, such works eschew learning altogether. We call them {\em fixed representation\/} methods.
The Family of Spectral Distances (FSGD)~\cite{verma2017} produces a high-dimensional sparse representation as a histogram on the dense biharmonic graph kernel; however, FGSD does not capture graph features at different scales of resolution or graph sizes, and is also inapplicable to reasonably large graphs, due to its quadratic time complexity.

By contrast to the above, we devise a lightweight {\em self-learned\/} representation method, which is extracted directly from the graph structure and can be used across graph analysis tasks. %

\section{Problem statement}\label{sec:problem-statement}

An undirected graph is a pair $\graphexp{}$, where $\vertices = (v_1, \ldots, v_n), n = |\vertices|$ is the set of vertices and $\edges\subseteq (\vertices \times \vertices)$ the set of edges. %
Since the vertex set is isomorphic to $\{1,\dots,n\}$, we will henceforth use the latter notation.
We assume the graph is unweighted, yet our method readily applies to the weighted case.
A \emph{representation} is a function $\bb{\sig} : \graphcol \rightarrow (\mathbb{R}^N,\ell_2)$ from a graph $\graph$ in a collection of graphs $\graphcol$ to the $N$-dimensional space equipped with the Euclidean metric; the element $j$ of the representation is denoted as $\sigma_j(\graph)$.
Notably, once a graph's representation is computed, comparisons between representations (e.g., for retrieval or classification) is independent of graph size. 

The first and foremost property a representation has to satisfy is \emph{permutation-invariance}, implying that if two graphs have the same structure (i.e., are isomorphic) the distance between their representations is zero. In other words, we demand that for every graph $G$, the representation is invariant to every permutation $\bb{\pi}$ of the graph vertices, $\bb{\sigma} \circ \bb{\pi}(G) = \bb{\sigma}(G)$.
In the sequel, we propose representations based on the Laplacian spectrum, which are permutation-invariant by construction.

Another desirable property is \emph{scale\hyp{}adaptivity}, implying that the representation shall account for both local (edge and node) and global (community) graph features.
A global feature is such that cannot be captured by \emph{any} combination of features on nodes at distance $r < \diam(\graph) - 1$, where $\diam(\graph)$ is the diameter (longest shortest-path length) of $\graph$. %
Let the set of all subgraphs of $\graph$ be $\xi(G)=\{g \sqsubset G: \diam(g)<\diam(\graph)\}$.
We define scale\hyp{}adaptivity as the property of a representation $\bb{\sig}$ having at least one local feature (i.e., derived solely from information encoded in subgraphs $\xi(G)$), and at least one global feature (i.e., derived by strictly more than the information encoded in any $\xi(G)$).
Using local features only, a similarity measure would deem two graphs sharing local patterns to have near-zero distance although their global properties (e.g., page-rank features) may differ; in reverse, relying on global features only would miss local ones (e.g., degree distribution).

We construct a parametric family of graph representations $\bb{\sig}^{\bb{\theta}}: \graphcol \rightarrow (\mathbb{R}^N,\ell_2)$, with parameter set $\bb{\theta}$, such that $\bb{\sig}^{\bb{\theta}}$ captures global and local features to different extents, depending on $\bb{\theta}$. Further, we adapt $\bb{\theta}$ to fit a purpose by means of unsupervised self-learning. %

\section{Spectral graph representations}\label{sec:solution}

\begin{figure*}
    \begin{center}
        \resizebox{\textwidth}{!}{%
\begin{tikzpicture}
\foreach \nodename/\nc in {
    0/255.0, 1/140.69473266601562, 2/82.0641860961914, 3/60.42073059082031, 4/54.074283599853516, 5/38.2684326171875, 6/26.949665069580078, 7/25.374380111694336, 8/39.29713821411133, 9/48.753868103027344, 10/46.93553924560547, 11/43.515289306640625, 12/40.291439056396484, 13/41.39219665527344, 14/31.043472290039062, 15/33.621437072753906, 16/32.90702819824219, 17/30.439523696899414, 18/36.61711883544922, 19/34.755043029785156, 20/34.25773620605469, 21/30.76152992248535, 22/26.854549407958984, 23/24.28759765625, 24/24.523635864257812, 25/19.937105178833008, 26/14.99986743927002, 27/17.4927921295166, 28/14.734357833862305, 29/9.656566619873047, 30/10.910247802734375, 31/5.647429466247559, 32/6.766271591186523, 33/5.0971903800964355, 34/7.504331588745117, 35/7.257763385772705, 36/9.715706825256348, 37/7.775983810424805, 38/5.428762912750244, 39/7.616656303405762, 40/10.219182014465332, 41/11.28637409210205, 42/15.812628746032715, 43/9.628540992736816, 44/5.63515567779541, 45/12.97339916229248, 46/16.237510681152344, 47/18.76555824279785, 48/12.56855583190918, 49/17.192569732666016, 50/13.359703063964844, 51/12.536064147949219, 52/15.531754493713379, 53/12.862730979919434, 54/10.131325721740723, 55/6.616556644439697, 56/8.336302757263184, 57/9.355871200561523, 58/6.387488842010498, 59/12.406954765319824, 60/8.562568664550781, 61/8.420787811279297, 62/11.525269508361816, 63/8.20522689819336, 64/5.262474060058594, 65/7.071977615356445, 66/4.344838619232178, 67/3.308149814605713, 68/8.641087532043457, 69/5.912705421447754, 70/6.801898956298828, 71/3.949153184890747, 72/7.0606489181518555, 73/5.455047130584717, 74/2.8383073806762695, 75/7.444589614868164, 76/4.665779113769531, 77/0.0, 78/8.87369441986084, 79/2.67049503326416, 80/2.062765121459961, 81/5.0475382804870605, 82/9.78361701965332, 83/6.091056823730469, 84/5.2399420738220215, 85/10.652878761291504, 86/3.719924211502075, 87/6.840485572814941, 88/6.844962120056152, 89/9.808223724365234, 90/7.139570713043213, 91/6.89497709274292, 92/5.128045558929443, 93/4.453674793243408, 94/4.503438472747803, 95/1.9134162664413452, 96/7.120339393615723, 97/8.306734085083008, 98/1.1176283359527588, 99/3.3779098987579346, 100/5.784331321716309, 101/4.600020408630371, 102/7.678250312805176, 103/6.08870792388916, 104/8.262598991394043, 105/0.3933887481689453, 106/7.223090171813965, 107/5.317452430725098, 108/4.0535383224487305, 109/4.403509616851807, 110/7.40330696105957, 111/9.920244216918945, 112/6.836205959320068, 113/6.573505878448486, 114/5.2350993156433105, 115/9.956456184387207, 116/9.916075706481934, 117/9.478943824768066, 118/6.843574523925781, 119/12.849778175354004, 120/11.854168891906738, 121/9.28874397277832, 122/5.674314975738525, 123/11.38005256652832, 124/7.27322244644165, 125/11.606537818908691, 126/9.494003295898438, 127/10.556427955627441, 128/3.433304786682129, 129/3.0599560737609863, 130/12.922528266906738, 131/11.960953712463379, 132/12.79050350189209, 133/14.70938491821289, 134/17.39886474609375, 135/15.413674354553223, 136/21.142614364624023, 137/17.363338470458984, 138/17.73973274230957, 139/14.65336799621582, 140/9.847204208374023, 141/10.892634391784668, 142/2.6774163246154785, 143/5.818943977355957, 144/6.539284706115723, 145/8.133153915405273, 146/12.006948471069336, 147/5.122926712036133, 148/5.657659530639648, 149/12.27226734161377, 150/8.709253311157227, 151/10.86492919921875, 152/6.746592044830322, 153/10.611680030822754, 154/11.878748893737793, 155/10.871235847473145, 156/9.795204162597656, 157/11.385327339172363, 158/10.679810523986816, 159/11.066987037658691, 160/8.03824520111084, 161/9.332510948181152, 162/7.3041605949401855, 163/3.3597588539123535, 164/5.651619911193848, 165/5.231094837188721, 166/6.718433380126953, 167/8.549400329589844, 168/8.128771781921387, 169/6.321612358093262, 170/3.5965137481689453, 171/11.008781433105469, 172/9.514095306396484, 173/7.451562881469727, 174/5.881822109222412, 175/12.237284660339355, 176/9.241755485534668, 177/11.787271499633789, 178/9.955127716064453, 179/10.861831665039062, 180/12.69577407836914, 181/14.199840545654297, 182/10.121612548828125, 183/13.98440933227539, 184/15.208246231079102, 185/14.043020248413086, 186/14.470921516418457, 187/10.861568450927734, 188/11.517662048339844, 189/12.987319946289062, 190/11.003005981445312, 191/11.786698341369629, 192/10.135181427001953, 193/8.69678783416748, 194/10.340164184570312, 195/7.559503078460693, 196/16.39753532409668, 197/13.297801971435547, 198/15.692741394042969, 199/13.716092109680176, 200/15.531598091125488, 201/16.292213439941406, 202/13.650492668151855, 203/11.239869117736816, 204/14.222284317016602, 205/16.07558822631836, 206/20.92679786682129, 207/14.307075500488281, 208/15.393403053283691, 209/16.223731994628906, 210/19.237802505493164, 211/20.422382354736328, 212/19.247507095336914, 213/20.136404037475586, 214/20.751726150512695, 215/21.021039962768555, 216/18.71649742126465, 217/18.716651916503906, 218/19.133590698242188, 219/25.343889236450195, 220/22.228961944580078, 221/25.020410537719727, 222/23.337650299072266, 223/24.69081687927246, 224/19.557851791381836, 225/22.721946716308594, 226/18.998010635375977, 227/22.42504119873047, 228/19.60906219482422, 229/22.005395889282227, 230/26.555076599121094, 231/20.339113235473633, 232/24.227630615234375, 233/27.206371307373047, 234/29.16282081604004, 235/28.951210021972656, 236/33.33449935913086, 237/27.39316749572754, 238/28.935590744018555, 239/31.895408630371094, 240/27.72040367126465, 241/34.28163528442383, 242/33.30851364135742, 243/32.19290542602539, 244/32.148658752441406, 245/37.35914611816406, 246/31.712509155273438, 247/35.2473258972168, 248/35.044498443603516, 249/33.78395080566406, 250/38.8890495300293, 251/37.38157272338867, 252/38.058349609375, 253/46.98081970214844, 254/34.22712707519531, 255/34.26390838623047
    }
{
    \fill[fill=cycle5!\nc!white,draw=cycle5!\nc!white] (\nodename,1) rectangle +(1,10);
}
\foreach \nodename/\nc in {
    0/255.0, 1/35.743133544921875, 2/16.935672760009766, 3/22.80412483215332, 4/8.179168701171875, 5/16.006284713745117, 6/28.268253326416016, 7/31.502002716064453, 8/16.868122100830078, 9/1.5948728322982788, 10/0.0, 11/0.5583621263504028, 12/7.638777732849121, 13/13.725851058959961, 14/28.162668228149414, 15/29.980518341064453, 16/26.14207649230957, 17/36.205162048339844, 18/36.20734786987305, 19/40.92714309692383, 20/42.174530029296875, 21/46.23653793334961, 22/50.80989456176758, 23/56.097938537597656, 24/56.478694915771484, 25/62.46533203125, 26/69.17850494384766, 27/69.62371826171875, 28/71.265869140625, 29/75.61278533935547, 30/80.00983428955078, 31/85.28060150146484, 32/78.99122619628906, 33/80.71180725097656, 34/84.51554870605469, 35/82.50920104980469, 36/82.10575866699219, 37/86.68531799316406, 38/86.75008392333984, 39/86.96148681640625, 40/84.50897216796875, 41/86.06059265136719, 42/86.91661071777344, 43/88.66349029541016, 44/92.00687408447266, 45/84.37580108642578, 46/81.83684539794922, 47/84.43572998046875, 48/88.97892761230469, 49/86.49027252197266, 50/91.07389068603516, 51/90.00857543945312, 52/86.77752685546875, 53/91.51349639892578, 54/95.09375, 55/94.4412612915039, 56/93.71210479736328, 57/98.29859161376953, 58/93.20845794677734, 59/93.7524642944336, 60/93.92234802246094, 61/93.03330993652344, 62/89.7311019897461, 63/94.95964813232422, 64/94.4917984008789, 65/98.57801055908203, 66/97.81500244140625, 67/99.2408676147461, 68/95.46141052246094, 69/103.74160766601562, 70/101.74991607666016, 71/101.02616119384766, 72/99.17988586425781, 73/101.50772094726562, 74/99.5074462890625, 75/96.9809799194336, 76/101.62189483642578, 77/104.6922378540039, 78/98.86695861816406, 79/104.37782287597656, 80/104.51485443115234, 81/103.78855895996094, 82/97.6502914428711, 83/98.64445495605469, 84/97.45087432861328, 85/94.836669921875, 86/101.58149719238281, 87/102.09637451171875, 88/96.25484466552734, 89/93.37422180175781, 90/97.19562530517578, 91/96.19991302490234, 92/98.85488891601562, 93/100.5578842163086, 94/95.42798614501953, 95/106.33116912841797, 96/99.52948760986328, 97/95.90299987792969, 98/103.16226959228516, 99/103.77701568603516, 100/101.17505645751953, 101/99.501953125, 102/101.40802764892578, 103/100.42205810546875, 104/94.99085998535156, 105/102.34320831298828, 106/99.47124481201172, 107/95.5328140258789, 108/97.701171875, 109/96.8474349975586, 110/93.3634033203125, 111/96.80211639404297, 112/100.21959686279297, 113/94.09676361083984, 114/92.85452270507812, 115/97.3079605102539, 116/93.47428131103516, 117/95.19390869140625, 118/90.27545928955078, 119/85.31723022460938, 120/87.48416137695312, 121/83.18772888183594, 122/91.09613800048828, 123/87.74858093261719, 124/91.9901123046875, 125/88.10211944580078, 126/85.54975891113281, 127/87.15803527832031, 128/94.6717300415039, 129/98.70299530029297, 130/87.01448059082031, 131/92.38260650634766, 132/94.60160064697266, 133/90.32825469970703, 134/89.56859588623047, 135/94.1401596069336, 136/86.2041244506836, 137/89.31768798828125, 138/92.19287109375, 139/95.9538345336914, 140/99.05906677246094, 141/101.2275390625, 142/107.22808074951172, 143/105.40025329589844, 144/100.13042449951172, 145/95.65705871582031, 146/94.25462341308594, 147/99.51250457763672, 148/94.5199966430664, 149/90.25871276855469, 150/96.83604431152344, 151/92.81060028076172, 152/95.06400299072266, 153/91.51243591308594, 154/84.66326904296875, 155/86.4085464477539, 156/89.06732177734375, 157/78.6976089477539, 158/82.37223052978516, 159/84.5805892944336, 160/90.69024658203125, 161/87.0551986694336, 162/96.00782012939453, 163/99.70184326171875, 164/94.8504409790039, 165/102.07593536376953, 166/92.12519073486328, 167/96.02043151855469, 168/90.5928955078125, 169/96.15066528320312, 170/91.40514373779297, 171/82.75860595703125, 172/84.22050476074219, 173/82.67291259765625, 174/81.18453216552734, 175/74.09307098388672, 176/80.05998229980469, 177/76.86585998535156, 178/79.62964630126953, 179/75.32678985595703, 180/75.7950439453125, 181/72.46219635009766, 182/75.06094360351562, 183/73.26178741455078, 184/65.55560302734375, 185/71.2815933227539, 186/76.79156494140625, 187/83.1004409790039, 188/76.05619812011719, 189/76.56021118164062, 190/80.47314453125, 191/80.15863037109375, 192/87.01219940185547, 193/86.2873764038086, 194/75.52206420898438, 195/83.64002227783203, 196/75.87226104736328, 197/79.20382690429688, 198/78.9444580078125, 199/79.83714294433594, 200/74.78048706054688, 201/73.58264923095703, 202/76.14170837402344, 203/78.55126190185547, 204/74.58222198486328, 205/77.99566650390625, 206/72.46477508544922, 207/79.13910675048828, 208/80.0558090209961, 209/75.64363098144531, 210/74.31262969970703, 211/68.44711303710938, 212/69.013916015625, 213/74.78135681152344, 214/72.2328872680664, 215/70.47388458251953, 216/70.31873321533203, 217/69.37332153320312, 218/65.20822143554688, 219/65.29048919677734, 220/64.1860580444336, 221/61.13922882080078, 222/59.68727493286133, 223/56.84882736206055, 224/60.804710388183594, 225/60.56058120727539, 226/56.99009704589844, 227/50.98090362548828, 228/55.4560432434082, 229/49.053733825683594, 230/45.71360397338867, 231/49.47564697265625, 232/48.39054489135742, 233/44.61071014404297, 234/41.64786148071289, 235/37.93038558959961, 236/36.511070251464844, 237/40.9649543762207, 238/35.14259719848633, 239/37.35103988647461, 240/39.55159378051758, 241/32.459991455078125, 242/34.5109977722168, 243/35.386043548583984, 244/37.40918731689453, 245/29.49832534790039, 246/32.43675994873047, 247/35.090003967285156, 248/31.72864532470703, 249/32.98310089111328, 250/33.75033187866211, 251/40.1282958984375, 252/40.33647155761719, 253/34.40381622314453, 254/55.036434173583984, 255/76.18173217773438
    }
{
    \fill[fill=cycle5!\nc!white,draw=cycle5!\nc!white] (\nodename,-13) rectangle +(1,10);
}
\draw[ultra thick] (0, 1) rectangle (256, 11);
\draw[ultra thick] (0, -13) rectangle (256, -3);
\end{tikzpicture} %
}%
    \end{center}
    \caption{Different regions of the spectrum have a different impact on the classifier co-trained with the \sgr. The color map shows the gradient magnitude of the classifier output with respect to the input spectrum visualized in increasing order from left to right, averaged on $600$ graphs. Top: SBM; bottom: Erd\H{o}s-R\'{e}nyi.}\label{fig:saliency}
\end{figure*}
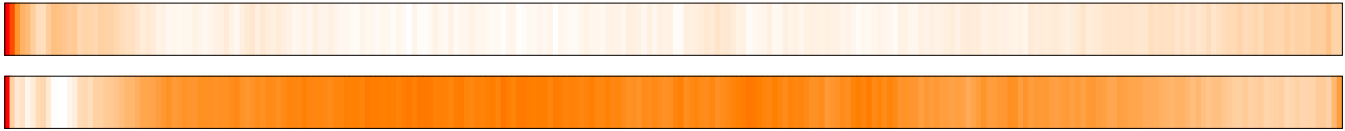

The \emph{adjacency matrix} of a graph $\graph$ is a $n\times n$ matrix $\adj$ having $A_{ij} \!=\! 1$ if $(i,j) \in \edges$ and $A_{ij}\! =\! 0$ otherwise. The \emph{normalized graph Laplacian} is defined as the matrix $\lapl\! =\!  \bb{I}\! -\! \bb{D}^{-\frac{1}{2}}\bb{A}\bb{D}^{-\frac{1}{2}}$, where $\bb{D}$ is the diagonal matrix with the degree of node $i$ as entry $D_{ii}$, i.e, $D_{ii} = \sum_{j = 1}^n A_{ij}$.
Since the Laplacian is a symmetric matrix, its eigenvectors $\bb{\phi}_1, \ldots, \bb{\phi}_n$, are real and orthonormal.
Thus, it is factorized as $\lapl = \bb{\Phi}\bb{\Lambda}\bb{\Phi}^\top$, where $\Lambda$ is a diagonal matrix with the sorted eigenvalues $\lambda_1 \le \ldots \le \lambda_n$, and $\Phi$ is the orthonormal matrix $\bb{\Phi} = (\bb{\phi}_1,  \dots, \bb{\phi}_n)$ having the corresponding eigenvectors as its columns.
Belkin~and~Niyogi~\cite{belkin2007} showed that the eigenvectors of the normalized Laplacian of a point cloud graph converge to the eigenfunction of the Laplace-Beltrami operator~\cite{berger2012panoramic} on the underlying Riemannian manifold.

The set of eigenvalues $\{\lambda_1, \ldots, \lambda_n\}$ is called the \emph{spectrum} of a graph and is bounded in the range $\lambda_i \in [0,2]$. 
Being an algebraic invariant of the Laplacian, its spectrum is independent of the choice of the basis, which, in particular means that it is permutation-invariant. 

\subsection{Heat propagation}
A useful metaphor for studying the graph structure at different scales is that of a system of connected heat-conducting rods corresponding to the graph edges. Heat propagation in such a system is governed by the
\emph{heat equation} associated with the Laplacian, 
\begin{equation}\label{eq:heat-diffeq}
    \frac{\partial \bb{u}_t}{\partial t} = -\lapl \bb{u}_t,
\end{equation}
\noindent where $\bb{u}_t$ is a scalar field on the graph vertices representing the amount of heat at each vertex at time $t$.
The solution to the heat equation provides the heat at each vertex at time $t$, when the initial heat $\bb{u}_0$ is initialized with a fixed value on one of the vertices.
It has a closed-form fundamental solution in the form of the $n\times n$ \emph{heat kernel} matrix,
\begin{equation}\label{eq:heat-solution}
    \hk = e^{-t\lapl} = \bb{\Phi}{}e^{-t\bb{\Lambda}}\bb{\Phi}^\top = \sum_{k} e^{-t \lambda_k} \bb{\phi}_k \bb{\phi}_k^\top,
\end{equation}
where ${(\hk)}_{ij}$ represents the amount of heat transferred from vertex $i$ to vertex $j$ in time $t$. 
The diagonal entries of $\hk$ are called the auto-diffusivity function, representing the amount of heat remaining at each graph vertex after time $t$. This auto-diffusivity function is affected by increasingly global structures of the graph's topology as the time parameter $t$ grows.

The sum of the auto-diffusivity function, known as the \emph{heat trace}
\begin{equation}\label{eq:heat-trace}
    h_t = \tr(\hk) = \sum_{j}{e^{-t\lambda_j}}
\end{equation}
is an algebraic invariant of the heat kernel and can be, therefore, expressed only in terms of the invariant graph spectrum.

Theoretical results by M{\'e}moli~\cite{memoli2011} subscribe the expressiveness of heat traces, suggesting a spectral definition of the Gromov\hyp{}Wasserstein distance between Riemannian manifolds based on matching the heat kernels at all scales. In what follows, we briefly review this construction, adapting it to graphs.  
Let us set the cost of matching a pair of vertices $(i,i')$ in a graph $G_1$ to a pair of points $(j,j')$ in a graph $G_2$ at a scale $t$ to be the discrepancy of the corresponding heat kernels,
$$
\bb{\Gamma}_t(i,j,i',j') = e^{-2(t+t^{-1})}\, | (\bb{H}_t^{G_1})_{ii'} - (\bb{H}_t^{G_2})_{jj'} |,
$$
where the factor $e^{-2(t+t^{-1})}$ %
scales the kernels.
A distance between the graphs can then be defined in terms of the minimal measure coupling 
$$
d^2(G_1,G_2) = \min_{\bb{M}} \sup_{t>0} \| \bb{\Gamma}_t \|_{\ell^2(\bb{M} \times \bb{M})}^2,
$$
where the minimum is sought over all doubly-stochastic matrices $\bb{M}$ representing a discrete measure on $G_1 \times G_2$ that marginalizes to the uniform measures on $G_1$ and $G_2$. This distance can be thought of as a ``soft'' version of the standard graph edit distance and has the useful property that $d(G_1,G_2) = 0$ iff $G_1$ and $G_2$ are isomorphic. 

M{\'e}moli~\cite{memoli2011} showed that the spectral Gromov-Wasserstein distance can be lower bounded by
$$
d(G_1,G_2)  \ge \sup_{t>0} e^{-2(t+t^{-1})} \, | h^{G_1}_t - h^{G_2}_t |,
$$
which is the scaled $L_\infty$ distance between heat traces of the graphs. 

\subsection{Learned spectral representations} 

The heat traces can be viewed as a nonlinear transformation of the graph spectrum of the form 
$\sum_{k} f_t(\lambda_k)$ 
with $f_t(\lambda) = e^{-t\lambda}$. Sampling the time parameter on some grid $\{t_1,\dots,t_N\}$ yields the following $N$-dimensional representation of the graph: 
$$
\bb{\sigma} = \left( \sum_{k} f_{t_1} (\lambda_k), \dots, \sum_{k} f_{t_N} (\lambda_k) \right).
$$

\begin{table*}[h!]
\setlength{\aboverulesep}{0pt}
\setlength{\belowrulesep}{0pt}
\begin{center}
{
\small
\newcolumntype{C}{>{\centering\arraybackslash}X}
\begin{tabularx}{\textwidth}{XCCCC|CCC|C} %
\multicolumn{1}{C}{} & \multicolumn{4}{c}{\em Kernels\/} & \multicolumn{3}{c}{\em Fixed Representations\/} & \multicolumn{1}{c}{\em Self-sup. Repr.} \\
\cmidrule(lr){2-5} \cmidrule(lr){6-8} \cmidrule(lr){9-9}
\emph{dataset} & \spg & \gk & \wl & \mlg & \netsimile & \fgsd & $\Lambda$ & \cellcolor{cycle2!10}{\sgr} \\ 
\midrule
D\&D & >1D & 73.39 & 68.27 & >1D 			& 70.02 & 64.88 & 64.54 & \win 76.12 \\
ENZYMES & 22.57 & 19.11 & 25.11 & 31.40 	& 28.06 & 28.85 & 25.28 & \win 33.67 \\
MUTAG & 80.30 & 80.76 & 81.16 & 86.54 		& 83.66 & 85.23 & 82.07 & \win 86.97 \\
PROTEINS & 72.04 & 71.48 & 72.33 & 73.10 	& 70.59 & 63.27 & 71.32 & \win 73.83 \\
\midrule
COLLAB & >1D & >1D & \win 78.52 & >1D & 74.26 & 70.66 & 66.15 & 71.98 \\
IMDB-B & 57.10 & 61.79 & \win 72.26 & 59.18 & 70.96 & 69.20 & 63.16 & 70.38 \\
IMDB-M & 39.39 & 39.80 & \win 50.75 & 34.31 & 46.80 		& 48.88 & 41.14 & 47.97 \\
\mbox{REDDIT-B} & >1D & 72.30 & 71.97 & >1D & 86.84 & 87.12 & 76.25 & \win 87.45 \\
\mbox{REDDIT-M-5k} & >1D & 23.39 & 48.57 & >1D & 44.96 & 48.51 & 48.02 & \win 53.22 \\
\bottomrule
\end{tabularx}
}
\end{center}
\caption{Graph classification accuracy on bio-chemical (top) and social (bottom) graph collections. Best results are highlighted.}\label{tab:classification-accuracy-new}
\end{table*} 
We propose to {\em extend this view\/} to a more general parametric family of spectrum transformations. Given a graph $G$ with $n$ vertices, we first compute its spectrum or a part thereof $\{ \lambda_k \}$, and interpolate it producing $\lambda(x)$ on the interval $[0,1]$ such that $\lambda(k/n) = \lambda_k$. The spectrum is then sampled on a fixed grid $(x_1,\dots,x_M)$ with $M$ points, producing an $M$-dimensional vector $\bb{\tilde{\lambda}}$ with the entries $\tilde{\lambda}_k = \lambda(x_k)$. Note that $\bb{\tilde{\lambda}}$ is insensitive to a graph's size and invariant to the ordering of its vertices. 

The interpolated and sampled spectrum $\bb{\tilde{\lambda}}$ undergoes next a parametric non-linear transformation implemented as a single-layered perceptron,
$$
\bb{\sigma} = \psi(\bb{W} \bb{\tilde{\lambda}} + \bb{b})
$$
where $\bb{W}$ is an $N \times M$ weight matrix, $\bb{b}$ is an $N$-dimensional bias vector, and $\psi$ is an element-wise SeLU non-linearity~\cite{klambauer2017self}. The resulting $N$-dimensional spectral graph representation (\ours) is parameterized by $\bb{\theta} = (\bb{W},\bb{b})$.

We propose a regime to train this representation. To obtain a representation capturing predominantly the \emph{global} structure of the graph (manifested in the lower part of the spectrum), we co-train $\bb{\sigma}$ jointly with a binary classifier attempting to distinguish between Erd\H{o}s-R\'{e}nyi random graphs and stochastic block model~\cite{karrer2011} graphs of various degrees and sizes, which have very different community structures. The binary classifier is embodied as a single linear layer on top of the output of $\bb{\sigma}$ followed by softmax, and is trained using the regular cross-entropy loss.

The classifier is tossed away, leaving an appropriately trained graph representation. This approach is inspired by the versatility of image embeddings obtained from deep neural networks trained on visual recognition tasks. We henceforth denote the representation as \sgr. Figure~\ref{fig:saliency} depicts the saliency map for the interpolated spectrum. Perhaps surprisingly, the neural network leaned to utilize not only the global information, but also very local part of the spectrum.

Full eigendecomposition takes $\bigO(n^3)$ time and $\bigO(n^2)$ space. While for graphs with $\Delta(G) \ll n$ the sparse structure of the Laplacian allows to reduce the complexity to $\bigO(n^2)$, it is still prohibitive for large graphs. Instead, we compute $k \ll n$ top and bottom eigenvalues, and use interpolation in between. This reduces complexity to $\bigO(n^2k)$ in the general case and to $\bigO(nk)$ in the case of bounded degree graphs. %
\section{Experiments}\label{sec:experiments}

We evaluate \thiswork on classification and clustering tasks on a variety of real graph collections. We compare against state-of-the-art kernels and graph representation methods, in terms of accuracy and running time. In order to ensure experimental repeatability we provide data, parameters, and source code\footnote{Will be available upon publication}.

\mpara{Experimental setup.}
We ran experiments on a 20-core Intel Xeon CPU E5-2640v4, 3.20GHz machine with 256Gb RAM\@. 
Unless otherwise stated, we repeat each experiment $100$ times and report the average across all trials.
\thiswork interpolates the spectrum of the normalized Laplacian of each graph in the collection through cubic splines; we use $256$ values uniformly sampled in the interpolated spectrum. 

We compare \thiswork{} against representative graph kernel methods: the Shortest-Path (\spg)~\cite{borgwardt2005} kernel, the Graphlet kernel (\gk)~\cite{shervashidze2009}, the Weisfeiler-Lehman kernel (\wl), and the state-of-the-art Multiscale Laplacian Graph kernel (\mlg)~\cite{kondor2016}, using default parameters for each method. 
We also compare \thiswork against \netsimile~\cite{berlingerio2013} and \fgsd~\cite{verma2017} graph representations.
We additionally report the results of a na\"ive baseline spectral representation ($\Lambda$) that represents the graphs with a $256$-dimensional vector sampled uniformly from a cubic spline-interpolated~\cite{dierckx1995curve} spectrum of the normalized Laplacian. 

\mpara{Datasets.}
We use $9$ graph collections from the standard benchmark for Graph Kernels~\cite{KKMMN2016}. 
Such collections describe either social interactions (e.g., REDDIT-B from messages in the Reddit platform) or biological connections (e.g., protein-protein interactions in PROTEINS). 
The number of graphs in each collection varies from $200$ (MUTAG) to $5000$ (REDDIT-M-5k), while the average graph size varies from $18$ (MUTAG) to $500$ (REDDIT).

\subsection{Classification}

In our classification experiment, on each of the datasets we randomly select 80\% of the data for training, and 20\% for testing. 
We train an SVM using LibSVM~\cite{chang2011libsvm} with default parameter $C{=}1$ and each kernel. 
For all graph representations, including \sgr, we train a logistic regression classifier with default $C{=}1$ and $L_2$ regularization. 
Table~\ref{tab:classification-accuracy-new} reports the classification accuracy averaged over $100$ runs.

Our method attains good quality in almost all datasets except for IMDB datasets, for which \fgsd outperforms \thiswork.
Due to the small average graph size and density of these datasets, the task becomes harder for our self-supervised approach that relies on local and global graph structures.
At the same time, while state-of-the-art kernels (MLG) outperform \thiswork, they fail to deliver results on medium and large collections in less than one day.  %

\section{Conclusions}\label{sec:conclusions}

We introduced \thiswork, a lightweight and concise graph representation that is self-learned by means of a single-layer perceptron over a collection of synthetically generated graphs. In particular, \ours learns a single-layer perceptron encoding global and local graph properties as nonlinear transformations of the graphs' Laplacian spectra; thus, it can adapt to a multitude of analytical tasks and application domains. Through extensive experimentation, we established that \ours achieves accuracy matching (or negligibly below) that of the most computationally demanding kernel methods on graph classification and clustering. In the future, we want to investigate more advanced architectures, and ways to incorporate both node and edge labels into the learning task.

\bibliographystyle{ACM-Reference-Format}
\bibliography{bibliography} 
\end{document}